# Title Page

**Title: Precision in the Face of Noise - Lessons from Kahneman, Siboney, and Sunstein for Radiation Oncology**

**Authors**: Kareem A. Wahid, PhD[1,2], Clifton D. Fuller, MD, PhD[2], David Fuentes, PhD[1]

[1] Department of Imaging Physics, The University of Texas MD Anderson Cancer Center, Houston, TX, USA.

[2] Department of Radiation Oncology, The University of Texas MD Anderson Cancer Center, Houston, TX, USA.

**Contact:** kawahid@mdanderson.org.

**Funding Statement:** KAW was supported by an Image Guided Cancer Therapy (IGCT) T32 Training Program Fellowship from T32CA261856. CDF was supported by P30CA016672. DF was supported by R01CA195524.

**Conflicts of Interest:** KAW serves as an Editorial Board Member for Physics and Imaging in Radiation Oncology. CDF has received travel, speaker honoraria, and/or registration fee waivers unrelated to this project from Siemens Healthineers/Varian, Elekta AB, Philips Medical Systems,The American Association for Physicists in Medicine, The American Society for Clinical Oncology, The Royal Australian and New Zealand College of Radiologists, Australian & New Zealand Head and Neck Society, The American Society for Radiation Oncology, The Radiological Society of North America, and The European Society for Radiation Oncology.

**Declaration of generative AI and AI-assisted technologies in the writing process:** During the preparation of this work, the authors used a combination of Claude 3.5 Sonnet and ChatGPT 4o to improve the grammatical accuracy and semantic structure of portions of the text. After using these tools, the authors reviewed and edited the content as needed and take full responsibility for the content of the publication.

# Main Text

## Introduction

In their thought-provoking book *Noise: A Flaw in Human Judgment* [1], Kahneman, Sibony, and Sunstein shed light on a hidden threat to effective decision-making: the random variability in professional judgments (i.e., noise). Drawing from fields as diverse as law, economics, and forensic science, the authors offer valuable insights applicable to any domain where human judgment plays a central role in decision-making. For radiation oncologists, the concept of noise is particularly relevant as daily practice demands complex decisions under uncertainty. Radiation oncologists routinely choose between treatment modalities with competing advantages, delineate elusive tumor boundaries, and balance intricate dose constraint trade-offs for organs at risk. Each of these potentially noisy judgments carries significant weight, directly impacting patient clinical outcomes. A patient's ultimate outcome can feel like a "lottery", shaped by the particular physician they see and the underlying noise that influences their clinical decision-making process. Understanding and mitigating noise could lead to more consistent decision-making, enhanced treatment precision, and ultimately, improved patient care.

## Key Terminology in Understanding Noise

Kahneman, Sibony, and Sunstein describe judgment as a measurement where the human mind serves as the instrument, with overall error comprising two components: statistical bias and system noise. Bias represents systematic errors in one direction (*average of errors*), while noise refers to unwanted variability in judgments that should ideally be identical (*variability of errors*). **Figure 1** illustrates this distinction using contouring as an exemplar case. The total error can be expressed as the sum of squared bias and noise (mean squared error = $bias^2 + noise^2$). System noise, the focus of this book, can be broken down into level noise — consistent differences between judges — and pattern noise — inconsistencies in handling specific cases, with the total system noise expressed as: $system\ noise^2 = level\ noise^2 + pattern\ noise^2$. In radiation oncology, level noise might manifest as one oncologist consistently contouring larger target volumes than another, while pattern noise could appear as inconsistencies in how oncologists respond to particular tumor subtypes. Occasion noise, a subset of pattern noise, stems from transient factors like fatigue or time pressure. Historically, bias has received more attention than noise, partly due to our familiarity with psychological biases and tendency towards causal thinking. Importantly, psychological biases can contribute to either statistical bias or system noise, depending on whether they are broadly shared or manifest differently across individuals. Noise, requiring statistical thinking, has been less discussed despite contributing equally to the error equation. This framework helps us understand the various sources of variability in professional judgments, including those in radiation oncology, and underscores the importance of recognizing and addressing noise as a hidden source of error.

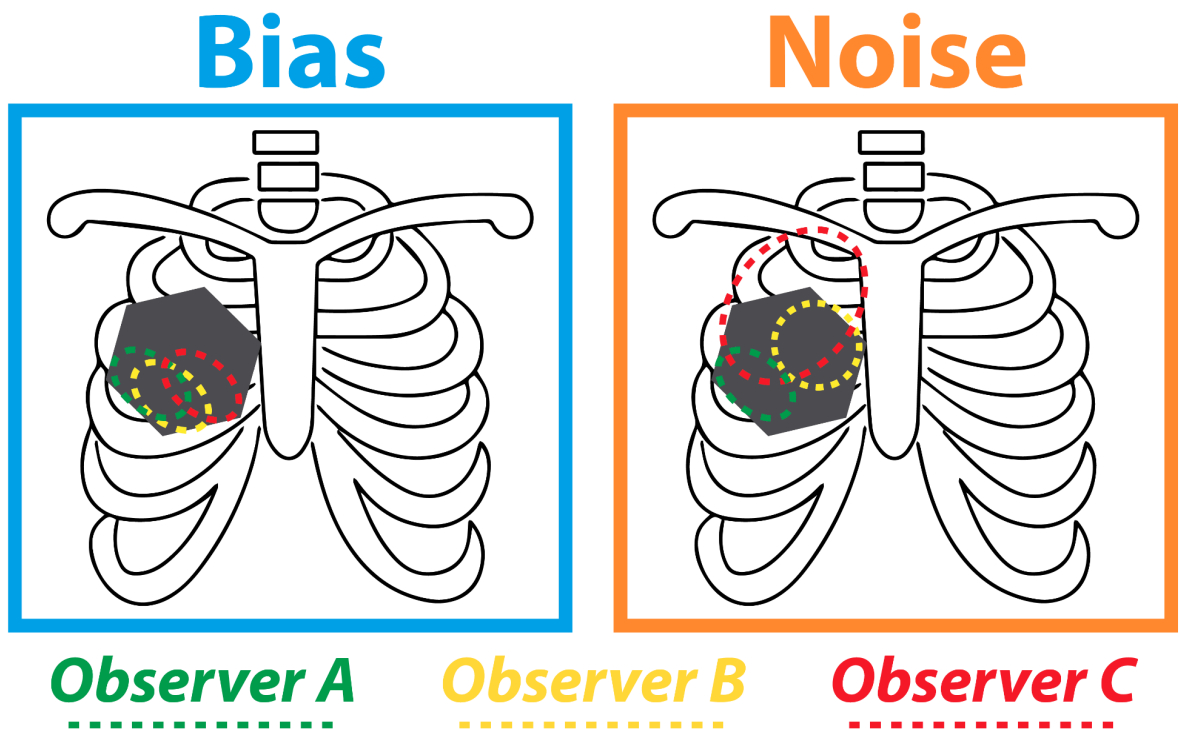

**Figure 1:** Illustration of bias and noise scenarios using radiation oncology contouring as a toy example. Assume the gray hexagon represents the true gross tumor volume. In the "bias" scenario, all three observers consistently under-contour the tumor in the same relative location, showing a systematic error. In the "noise" scenario, the observers' contours vary widely around the actual tumor, with some larger and some smaller, illustrating random variability in judgments. It's important to recognize that both bias and noise scenarios result in errors, and neither situation is inherently more desirable or "correct" than the other.

# Strategies for Identifying and Reducing Noise in Practice

The authors introduce two crucial concepts to identify and combat noise in decision-making processes. The first is the "noise audit", a rigorous method to quantify and analyze noise within an organization or system. This approach allows for the identification and measurement of unwanted variability in judgments, providing a baseline for improvement. Building on the previous examples, a relevant application of a potential noise audit in radiation oncology could involve regularly measuring departmental interobserver variability among clinicians in tumor contouring to establish reference standards for expected geometric variation. The second concept is "decision hygiene", a set of practices designed to reduce noise, analogous to how handwashing prevents the spread of unknown pathogens. Decision hygiene emphasizes the importance of preventive measures to minimize unwanted variability in judgments before they occur. In the context of radiation oncology, these concepts offer valuable frameworks for improving the consistency and quality of clinical decisions. By first conducting noise audits to understand the extent of the problem, and then implementing decision hygiene practices, radiation oncology departments can take proactive steps to enhance the reliability of their judgments and, ultimately, the quality of patient care.

## Historical Approaches to Reducing Noise in Radiation Oncology

Historically, multidisciplinary tumor boards [2] and peer review [3,4] have been key strategies in radiation oncology for combating noise. These collaborative forums bring together clinicians to discuss complex cases, review treatment strategies, and ensure consistency in patient care. These practices align with the book's decision hygiene concept of judgment aggregation, aiming to leverage the "wisdom of crowds". As the authors point out, "eliciting and aggregating judgments that are both independent and diverse will often be the easiest, cheapest, and most broadly applicable decision hygiene strategy". However, despite their widespread recognition within radiation oncology, it's worth noting that current tumor board and peer review implementations may not fully realize their noise-reduction potential. Simultaneous discussions, which often occur during these sessions, can lead to social influence in group decision-making, potentially propagating errors rather than fostering independent thought. While the authors don't specifically address tumor boards or peer review, their principles for effective aggregation of judgments can be applied to these contexts. For instance, an emphasis on having members independently record recommendations before group discussion could preserve the benefits of diverse expertise while mitigating groupthink and maintaining independence, though this approach would naturally need to be balanced against clinician time constraints. Adapting these key factors from the book's broader noise reduction strategies could enhance the effectiveness of these important radiation oncology practices.

Beyond the aforementioned collaborative approaches, radiation oncology has also developed strategies to address system noise in more technical aspects of treatment planning. Two areas where noise is particularly evident are contouring and treatment plan evaluation. As previously alluded to, contouring variability illustrates both level noise (consistent differences between oncologists) and pattern noise (inconsistencies in handling specific cases). To combat this, the field has developed detailed contouring guidelines [5], which provide standardized definitions for target volumes and organs at risk, often including visual atlases and specific instructions for different treatment sites and modalities. These guidelines aim to reduce level and pattern noise introduced by inter-observer variability in delineating structures critical for treatment planning. Similarly, to address the complexity of treatment plans and the subjective nature of plan evaluation, many institutions have implemented radiotherapy planning related checklists [6]. These checklists offer criteria for evaluating treatment plans, aligning with the authors' recommendations for decision hygiene and structured judgment processes. Collectively, these strategies have led to greater consistency in treatment delivery and enhanced patient outcomes. However, as with tumor boards and peer review, these structured approaches are not without limitations. Strict adherence to guidelines and checklists may sometimes constrain clinical judgment in unique cases. Moreover, the effectiveness of these tools depends on how they are implemented and updated to reflect evolving best practices. Balancing standardization with the need for flexibility in complex clinical scenarios remains an ongoing challenge in the field.

## The Future of Noise Reduction in Radiation Oncology

As we look to the future of radiation oncology, the insights from Kahneman, Sibony, and Sunstein provide valuable considerations for ongoing improvement in clinical decision-making. A promising approach to reducing noise in radiation oncology is the integration of artificial intelligence (AI) models, particularly in the form of auto-contouring and auto-planning tools [7]. As the authors highlight, even simple algorithmic models can often significantly reduce noise in decision-making processes, i.e., "the model of you beats you". However, these approaches must be balanced against over-reliance on AI, which could introduce new forms of bias and noise. To strike this balance, the field could benefit from cultivating "superforecaster" skills among clinicians. Superforecasters, as described in the book, are individuals who excel at making accurate predictions by combining probabilistic reasoning and "active open-mindedness" — leveraging an underlying Bayesian framework of continuously updating beliefs based on new information. Cultivating these cognitive skills in radiation oncology trainees could enhance clinicians' decision-making abilities while simultaneously reducing noise in treatment. Furthermore, integrating explicit uncertainty quantification in clinical practice — as is currently being explored with auto-contouring [8] — aligns well with the book's focus on understanding and managing uncertainty in decision-making. By formally acknowledging and accounting for the inherent uncertainties in target delineation and treatment planning, we can potentially develop more robust, noise-resistant protocols. Ultimately, the future of radiation oncology lies in thoughtfully integrating technological advancements with enhanced human judgment, creating a synergy that addresses the multifaceted challenge of noise in clinical practice.

## Conclusion

*Noise: A Flaw in Human Judgment* offers a novel perspective on improving decision-making in radiation oncology. While human judgment remains irreplaceable in our field, there is significant scope for its improvement. Embracing structured decision processes, cultivating probabilistic thinking, and judiciously integrating technology can help standardize care and reduce the "lottery" aspect of clinical decision-making. Radiation oncology departmental leaders might consider exploring noise audits and decision hygiene principles as potential avenues for enhancing clinical processes. While this editorial only scratches the surface of the book's rich content, we hope it inspires further exploration of these concepts within our field. As the authors aptly state, "Noise is an invisible enemy, and a victory against an invisible enemy can only be an invisible victory. But like physical health hygiene, decision hygiene is vital. After a successful operation, you like to believe that it is the surgeon's skill that saved your life—and it did, of course—but if the surgeon and all the personnel in the operating room had not washed their hands, you might be dead. There may not be much glory to be gained in hygiene, but there are results".